\documentstyle[twoside,fleqn,espcrc2,oldlfont]{article}

\newcommand{\AmS}{{\protect\the\textfont2
  A\kern-.1667em\lower.5ex\hbox{M}\kern-.125emS}}
\newcommand{\n}{\hspace*{-2.5mm}}
\newcommand{\simgt}{\rlap{\lower 3.5 pt \hbox{$\mathchar \sim$}} \raise 1pt
 \hbox {$>$}}
\newcommand{\simlt}{\rlap{\lower 3.5 pt \hbox{$\mathchar \sim$}} \raise 1pt
 \hbox {$<$}}

\title{Higher-order corrections to Higgs-boson decays\thanks{To
appear in the {\it Proceedings of the 1994 Zeuthen Workshop on Elementary
Particle Theory: Physics at LEP200 and Beyond}, Teupitz, Germany,
April 10--15, 1994, edited by J. Bl\"umlein and T. Riemann, Nucl.\ Phys.\ B
(Proceedings Supplements). }}
\author{Bernd A. Kniehl\\
II. Institut f\"ur Theoretische Physik,
Universit\"at Hamburg,\\
Luruper Chaussee 149, 22761 Hamburg, Germany}
\date{May 1994}
\begin{document}

\begin{abstract}
We review the status of radiative corrections to the decay rates of
the Standard-Model Higgs boson in the mass range accessible at LEP~2.
Specifically, we consider corrections of ${\cal O}(\alpha)$,
${\cal O}(\alpha_sG_Fm_t^2)$, and ${\cal O}(G_F^2M_H^4)$ to the
fermionic decay rates and two-loop QCD corrections to the hadronic
decay rates.

hep-ph/9405317
\end{abstract}

\maketitle

\section{Introduction}

One of the great puzzles of elementary particle physics today
is whether nature makes use of the Higgs mechanism of
spontaneous symmetry breaking to generate the observed particle masses.
The Higgs boson, $H$, is the missing link sought to verify
this concept in the Standard Model.
Many of the properties of the Higgs boson are fixed, e.g.,
its couplings to the gauge bosons,
$g_{VVH}=2^{5/4}G_F^{1/2}M_V^2\ (V=W,Z)$, and fermions,
$g_{f\bar fH}=2^{1/4}G_F^{1/2}m_f$,
and the vacuum expectation value,
$v=2^{-1/4}G_F^{-1/2}\approx246$~GeV.
However, its mass, $M_H$, and its self-couplings, which depend on $M_H$,
are essentially unspecified.

The failure of experiments at LEP~1 and SLC to detect the decay
$Z\rightarrow f\bar f H$ has ruled out the mass range $M_H\le63.8$~GeV at
the 95\% confidence level \cite{sch}.
At the other extreme, unitarity arguments in intermediate-boson scattering
at high energies \cite{dic} and considerations concerning the range of
validity of perturbation theory \cite{vel} establish an upper bound on
$M_H$ at $\left(8\pi\sqrt2/3G_F\right)^{1/2}\approx1$~TeV in a weakly
interacting Standard Model.

The Higgs-boson discovery potential of LEP~1 and SLC is almost exhausted
\cite{gro}.
Prior to the advent of the LHC, the Higgs-boson search will be restricted
to the lower mass range.
With LEP~2 it should be possible to find a Higgs boson with $M_H\le100$~GeV
when high energy and luminosity can be achieved \cite{gkw}.
A possible 4-TeV upgrade of the Tevatron might cover the $M_H$ range up to
120~GeV or so \cite{gun}.
At an $e^+e^-$ linear collider operating at 300~GeV,
50~fb$^{-1}$ luminosity and a $b$-tagging efficiency of 50\%
would be sufficient to detect a Higgs boson with $M_H\le150$~GeV in
the $\mu^+\mu^-b\bar b$ channel \cite{imh}.

Below the onset of the $W^+W^-$ threshold,
the Standard-Model Higgs boson is relatively long-lived,
with $\Gamma_H<100$~MeV, so that, to good approximation,
its production and decay processes may be treated independently.
The low-mass Higgs boson, with $M_H\le M_Z$, decays with more than
99\% probability into a fermion pair \cite{bar}.
With $M_H$ increasing, the $W^+W^-$ mode, with at least one
$W$ boson being off shell, gradually gains importance.
Its branching fraction surpasses that of the $\tau^+\tau^-$ mode at
$M_H\approx115$~GeV and that of the $b\bar b$ mode at $M_H\approx135$~GeV
\cite{bar}.
In the near future, however, Higgs-boson searches will rely mostly on the
$f\bar f$ modes.

Quantum corrections to Higgs-boson phenomenology have received
much attention in the literature; for a review, see Ref.~\cite{bak}.
The experimental relevance of radiative corrections to the $f\bar f$
branching fractions of the Higgs boson has been emphasized recently
in the context of a study dedicated to LEP~2 \cite{gkw}.
Techniques for the measurement of these branching fractions at a
$\sqrt s=500$~GeV $e^+e^-$ linear collider have been elaborated in
Ref.~\cite{hil}.

\section{Electroweak corrections to $\Gamma\left(H\to f\bar f\,\right)$}

In the Born approximation, the $f\bar f$ partial widths of the Higgs boson
are given by \cite{res}
\begin{equation}
\label{born}
\Gamma_0\left(H\to f\bar f\,\right)={N_cG_FM_Hm_f^2\over4\pi\sqrt2}
\left(1-{4m_f^2\over M_H^2}\right)^{3/2}\n,
\end{equation}
where $N_c=1$~(3) for lepton (quark) flavours.

\subsection{One-loop approximation}

The full one-loop electroweak corrections to Eq.~(\ref{born}) are now well
established \cite{fle,hff}.
They consist of an electromagnetic and a weak part, which are separately
finite and gauge independent.
They may be written as an overall factor,
$\left(1+(\alpha/\pi)Q_f^2\Delta_{\rm em}+\Delta_{\rm weak}\right)$.
For $M_H\to2m_f$, $\Delta_{\rm em}$ exhibits a threshold singularity with
a positive sign,
\begin{equation}
\Delta_{\rm em}=3\zeta(2)\left({1\over\beta}+\beta\right)-1
+{\cal O}\left(\beta^2\ln\beta\right),
\end{equation}
where $\beta=\sqrt{1-4m_f^2/M_H^2}$ is the fermion velocity in the
centre-of-mass frame.
This divergence is tamed by the overall threshold factor $\beta^3$ of
Eq.~(\ref{born}).
For $M_H\gg2m_f$, $\Delta_{\rm em}$ develops a large logarithm,
\begin{equation}
\label{delem}
\Delta_{\rm em}=-{3\over2}\ln{M_H^2\over m_f^2}+{9\over4}
+{\cal O}\left({m_f^2\over M_H^2}\ln{M_H^2\over m_f^2}\right).
\end{equation}

For $M_H\ll2M_W$, the weak part is well approximated by \cite{hff}
\begin{eqnarray}
\label{weak}
\Delta_{\rm weak}&\n=\n&{G_F\over8\pi^2\sqrt2}\left\{K_fm_t^2
+M_W^2\left({3\over s_w^2}\ln c_w^2-5\right)
\right.\nonumber\\&\n+\n&\left.
M_Z^2\left[{1\over2}-3\left(1-4s_w^2|Q_f|\right)^2\right]\right\},
\end{eqnarray}
where $c_w^2=1-s_w^2=M_W^2/M_Z^2$, $K_b=1$, and $K_f=7$ for all other
flavours, except for top.
The $t\bar t$ mode will not be probed experimentally anytime soon
and we shall not be concerned with it in the remainder of this presentation.
Equation~(\ref{weak}) has been obtained by putting $M_H=m_f=0$
($f\ne t$) in the expression for the full one-loop weak correction.
It provides a very good approximation for $f=\tau$ up to
$M_H\approx135$~GeV and for $f=b$ up to $M_H\approx70$~GeV,
the relative deviation from the full weak correction being less than 15\%
in each case.
 From Eq.~(\ref{weak}) it is evident that the dominant effect is due to virtual
top quarks.
In the case $f\ne b$, the $m_t$ dependence is carried solely by the
renormalizations of the wave function and the vacuum expectation value
of the Higgs field and is thus flavour independent.
These corrections are of the same nature as those considered in
Ref.~\cite{cha}.
For $f=b$, there are additional $m_t$ dependent contributions from the
$b\bar bH$ vertex correction and the $b$-quark wave-function renormalization.
Incidentally, they cancel almost completely the universal $m_t$ dependence.
It is amusing to observe that a similar situation has been encountered in the
context of the $Z\to f\bar f$ decays \cite{akh,bee}.
In summary, the universal virtual-top-quark term will constitute the
most important part of the weak one-loop corrections to Higgs-boson
decays in the near future.
In Sect.~2.2, we shall present the two-loop gluon correction to this
term.

It is interesting to consider the high-$M_H$ limit of $\Delta_{\rm weak}$.
The leading term is of ${\cal O}(G_FM_H^2)$ and flavour independent,
and arises from loops in which only physical and unphysical Higgs bosons
circulate.
It reads \cite{mve}
\begin{eqnarray}
\label{velt}
\Delta_{\rm weak}&\n=\n&{G_FM_H^2\over8\pi^2\sqrt2}
\left({13\over2}-\pi\sqrt3\right)
\nonumber\\
&\n\approx\n&11.1\%\left({M_H\over1{\rm TeV}}\right)^2.
\end{eqnarray}
It may be extracted conveniently in the framework of the Higgs-Goldstone
scalar theory.
As we shall see in Sect.~2.3, this formalism carries over to
${\cal O}(G_F^2M_H^4)$.

\subsection{Two-loop ${\cal O}(\alpha_sG_Fm_t^2)$ corrections}

The universal ${\cal O}(G_Fm_t^2)$ term of $\Delta_{\rm weak}$ resides
inside the combination
\begin{equation}
\label{delta}
\delta=-{\Pi_{WW}(0)\over M_W^2}-\Re e\Pi_{HH}^\prime\left(M_H^2\right),
\end{equation}
where $\Pi_{WW}$ and $\Pi_{HH}$ are the unrenormalized self-energies
of the $W$ and Higgs bosons, respectively.
The same is true of its QCD correction.

For $M_H<2m_t$ and $m_b=0$, the one-loop term reads \cite{hff}
\begin{eqnarray}
\label{delzero}
\delta_0&\n=\n&N_c{G_Fm_t^2\over2\pi^2\sqrt2}\left[\left(1+{1\over2r}\right)
\sqrt{{1\over r}-1}\arcsin\sqrt r
\right.\nonumber\\
&\n-\n&\left.\vphantom{\sqrt{{1\over r}-1}}
{1\over4}-{1\over2r}\right],
\end{eqnarray}
where $r=(M_H^2/4m_t^2)$.
In the same approximation, the two-loop term may be written as \cite{two}
\begin{eqnarray}
\label{delone}
\delta_1&\n=\n&{N_cC_F\over4}\,{G_Fm_t^2\over2\pi^2\sqrt2}\,{\alpha_s\over\pi}
\left(6\zeta(3)+2\zeta(2)-{19\over4}
\right.\nonumber\\
&\n-\n&\left.\vphantom{19\over4}
\Re eH_1^\prime(r)\right),
\end{eqnarray}
where $H_1$ has an expression in terms of dilogarithms and trilogarithms.
In the heavy-quark limit ($r\ll1$), one has \cite{two}
\begin{equation}
H_1^\prime(r)=6\zeta(3)+3\zeta(2)-{13\over4}+{122\over135}r+{\cal O}(r^2).
\end{equation}
Combining Eqs.~(\ref{delzero},\ref{delone}) and retaining only the leading
high-$m_t$ terms, one finds the QCD-corrected coefficients $K_f$ for $f\ne b$,
\begin{equation}
K_f=7-2\left({\pi\over3}+{3\over\pi}\right)\alpha_s
\approx7-4.004\,\alpha_s.
\end{equation}
We recover the notion that, in Electroweak Physics, the one-loop
${\cal O}\left(G_Fm_t^2\right)$ terms get screened by their QCD corrections.
The QCD correction to the shift in $\Gamma\left(H\to f\bar f\,\right)$ induced
by a pair of quarks with arbitrary masses may be found in Ref.~\cite{two}.

\subsection{Two-loop ${\cal O}(G_F^2M_H^4)$ corrections}

We shall now outline the derivation of the ${\cal O}(G_FM_H^2)$ and
${\cal O}(G_F^2M_H^4)$ corrections to
$\Gamma\left(H\rightarrow f\bar f\,\right)$
\cite{dur}.
In the limit of $M_H\gg M_W$, power counting and inspection of coupling
constants (in the 't~Hooft-Feynman gauge) reveal that we may concentrate on
those one- and two-loop diagrams which involve only the physical Higgs boson
and the longitudinal polarization states of the intermediate bosons.
The fermion mass and wave-function renormalizations do not receive
contributions in the orders considered.
Even though the process $H\rightarrow f\bar f$ does not involve external gauge
bosons, the fact that the corrections of interest arise solely from diagrams
containing only virtual Higgs and gauge bosons allows us to simplify the
calculation enormously by using the Goldstone-boson equivalence theorem
\cite{cor}.
In particular, the dominant contributions for $M_H\gg M_W$ can be calculated
by replacing the  propagators of the gauge bosons by the propagators of the
corresponding scalar Goldstone bosons, and setting the gauge coupling
and the gauge-boson masses to zero \cite{cor}.

Once the equivalence theorem has been invoked, all the information
necessary for the calculation may be obtained from the Lagrangian of
the Higgs or symmetry-breaking sector of the Standard Model.
This characterizes the kinematics and interactions of the Higgs boson, $H$,
and the Goldstone bosons, $w^\pm$ and $z$.
The latter remain massless and satisfy a residual SO(3) symmetry.
The boson masses and wave functions are renormalized according to the
usual on-mass-shell procedure.
Furthermore, we fix the physical vacuum expectation value and quartic coupling
by $v=2^{-1/4}G_F^{-1/2}$ and $\lambda=G_FM_H^2/\sqrt2$, respectively.
Straightforward algebra shows that the Higgs-fermion Yukawa coupling
receives a multiplicative correction of the form $(Z_H/Z_w)^{1/2}$,
where $Z_H$ and $Z_w$ are the wave-function renormalizations of
$H$ and $(w^\pm,z)$.
Thus, the leading high-$M_H$ corrections to the fermionic Higgs-boson decay
rates can be included by multiplying Eq.~(\ref{born}) with the overall factor
$Z_H/Z_w$, independently of the fermions involved.

The wave-function renormalizations $Z_H$ and $Z_w$ are defined as
\begin{eqnarray}
Z_w^{-1}&\n=\n&1-\left.{d\over dp^2}\Pi_w^0(p^2)\right|_{p^2=0},\nonumber\\
Z_H^{-1}&\n=\n&1-\left.{d\over dp^2}\Re e\Pi_H^0(p^2)\right|_{p^2=M_H^2},
\end{eqnarray}
where $\Pi_w^0$ and $\Pi_H^0$ are the self-energy functions for the bare
fields.
Using dimensional regularization, we find that $Z^{-1}$ can be written
in factored form,
\begin{eqnarray}
\label{wave}
Z_\sigma^{-1}&\n=\n&\left(1+a_\sigma\hat\lambda\xi^\epsilon
+b_\sigma\hat\lambda^2\xi^{2\epsilon}\right)
\left(1+{3\over\epsilon}\hat\lambda^2\xi^{2\epsilon}\right)\nonumber\\
&\n+\n&{\cal O}\left(\hat\lambda^3\xi^{3\epsilon}\right),
\qquad\sigma=w,H,
\end{eqnarray}
where $\hat\lambda=(\lambda/16\pi^2)$ and $\xi=(4\pi\mu^2/M_H^2)$.
Here $\mu$ is the arbitrary scale parameter used in the calculation to keep
$\lambda$ dimensionless in $D=4-2\epsilon$ dimensions.
The terms in Eq.~(\ref{wave}) which are singular for $\epsilon\rightarrow 0$
divide out exactly in the ratio $Z_H/Z_w$ as they must, since
$\Gamma\left(H\rightarrow f\bar f\,\right)$ is a physical quantity.
The ratio is independent of $\xi$ or $\mu$ in the physical limit
$\epsilon\rightarrow 0$, and is given in this limit by
\begin{equation}
\label{eqres}
{Z_H\over Z_w}={1+a_w\hat\lambda+b_w\hat\lambda^2\over
1+a_H\hat\lambda+b_H\hat\lambda^2},
\end{equation}
where $a_w=1$, $a_H=2\pi\sqrt3-12\approx-1.117$,
$b_w\approx-24.760$, and $b_H\approx265.764$ \cite{dur}.
Equation\ (\ref{eqres}), which naturally emerges from our formalism,
automatically resums one-particle-reducible Higgs-boson self-energy
diagrams in a way that conforms with the standard procedure
in $Z$-boson physics; see, e.g., Ref.~\cite{bee}.
However, we have no control of terms beyond
${\cal O}\left(\hat\lambda^2\right)$,
and are not aware of any physical organizing principle analogous to that
provided at high energies by the renormalization group,
which would allow us to select an optimum resummation scheme.
Expanding Eq.~(\ref{eqres}) and discarding terms beyond
${\cal O}\left(\hat\lambda^2\right)$, we obtain the alternative representation
\begin{eqnarray}
\label{eqexp}
{Z_H\over Z_w}&\n=\n&1+(a_w-a_H)\hat\lambda+(b_w-b_H-a_wa_H\nonumber\\
&&{}+a_H^2)\hat\lambda^2\\
&\n\approx\n&1+11.1\%\left({M_H\over1{\rm TeV}}\right)^2
-78.6\%\left({M_H\over1{\rm TeV}}\right)^4,\nonumber
\end{eqnarray}
which extends Eq.~(\ref{velt}) to two loops.

We are now in a position to explore the implications of our results.
In Fig.~\ref{figone}, we show the leading electroweak
corrections to $\Gamma\left(H\rightarrow f\bar f\,\right)$ in the one- and
two-loop approximations with and without resummation of
one-particle-reducible higher-order terms plotted as functions of $M_H$.
We shall concentrate first on the expanded expression in Eq.~(\ref{eqexp}).
While the ${\cal O}(G_FM_H^2)$ term (upper dotted line)
gives a modest increase in the rates, by 11.1\% at $M_H=1$~TeV,
the situation changes drastically when the two-loop term is included.
The importance of this term increases so rapidly with $M_H$ that
it already cancels the one-loop term completely for $M_H=375$~GeV.
By $M_H=530$~GeV, it has twice the magnitude of the one-loop term,
and the sum of one- and two-loop corrections (lower dotted line) reaches the
same magnitude as the one-loop corrections, but with a reversed sign.
The perturbation series for $\Gamma\left(H\rightarrow f\bar f\,\right)$
in powers of $\lambda$ or $G_FM_H^2$ clearly ceases to converge usefully,
if at all, for values of $M_H$ beyond about 400~GeV.
A Higgs boson with a mass larger than 400~GeV effectively becomes a strongly
interacting particle in the electroweak processes which contribute to the
correction, a very surprising result.
Conversely, $M_H$ must not exceed approximately 400~GeV if the standard
electroweak perturbation theory is to be predictive for the decays
$H\to f\bar f$.
Note that, for $M_H\simgt400$~GeV, one cannot use the usual unitarization
schemes invoked in studies of $W_L^\pm,Z_L,H$ scattering to restore the
predictiveness for the Higgs-boson width, as no unitarity violation occurs.

One might expect to improve the perturbative result in the upper $M_H$ range
somewhat by resumming the one-particle-reducible contributions to the
Higgs-boson wave-function renormalization according to Eq.\ (\ref{eqres}).
This leads to an insignificant increase of the one-loop correction
(upper solid line), while the negative effect of the two-loop correction
is appreciably lessened (lower solid line), i.e.,
the ratio of two- to one-loop corrections is rendered more favourable
theoretically.
However, in the mass range below $M_H=600$~GeV, this effect is too feeble
to change our conclusions concerning the breakdown of perturbation theory.
Moreover, the resummation of reducible one-loop terms in the perturbation
series does not yield a proper estimate for the size of the two-loop terms,
so that there is no reason to favour this approach in the present problem.

\section{QCD corrections to $\Gamma(H\to2j)$}

Both $H\to q\bar q$ and $H\to gg$ lead to two-jet final states.
These processes are relevant phenomenologically for the search for
the low- and intermediate-mass Higgs boson at $e^+e^-$ colliders,
while they are very difficult to discern from background reactions
at hadron colliders.
Both decay channels receive significant QCD corrections.

\subsection{Next-to-next-to-leading order corrections to
$\Gamma\left(H\to q\bar q\right)$}

The one-loop QCD correction to $\Gamma\left(H\to q\bar q\right)$ emerges
from one-loop QED correction, discussed in Sect.~2.1, by substituting
$\alpha_sC_F$, where $C_F=\left(N_c^2-1\right)/(2N_c)=4/3$, for
$\alpha Q_f^2$ \cite{bra}.
This corresponds to the on-shell scheme, where the quark pole mass, $M_q$,
is used as a basic parameter.
 From Eq.~(\ref{delem}) it is apparent that, for $M_q\ll M_H/2$, large
logarithmic corrections occur.
In general, they are of the form $(\alpha_s/\pi)^n\ln^m(M_H^2/M_q^2)$,
with $n\ge m$.
Appealing to the renormalization-group equation, these logarithms may be
absorbed completely into the running quark mass, $m_q(\mu)$, evaluated
at scale $\mu=M_H$.
In this way, these logarithms are resummed to all orders and the perturbation
expansion converges more rapidly.
This observation gives support to the notion that the $q\bar qH$ Yukawa
couplings are controlled by the running quark masses.

The values of $M_q$ may be estimated from QCD sum rules.
To obtain $m_q(M_H)$, one proceeds in two steps.
Firstly, one evaluates $m_q(M_q)$ from
\begin{equation}
{M_q\over m_q(M_q)}=1+C_F{\alpha_s(M_q)\over\pi}
+{\cal K}_q\left({\alpha_s(M_q)\over\pi}\right)^2,
\end{equation}
with \cite{gra}
\begin{equation}
{\cal K}_q\approx16.11-1.04\sum_{i=1}^{N_F-1}\left(1-{M_i\over M_q}\right),
\end{equation}
where the sum extents over all quark flavours with $M_i<M_q$.
Specifically, ${\cal K}_c=13.3$ and ${\cal K}_b=12.4$ \cite{gra}.
Secondly, one determines $m_q(M_H)$ via the scaling law
\begin{equation}
m_q(M_H)=m_q(M_q){c_q(\alpha_s(M_H)/\pi)\over c_q(\alpha_s(M_q)/\pi)},
\end{equation}
where
\begin{eqnarray}
c_c(x)&\n=\n&\left({25\over6}x\right)^{12/25}(1+1.014x+1.389x^2),\nonumber\\
c_b(x)&\n=\n&\left({23\over6}x\right)^{12/23}(1+1.175x+1.501x^2).
\end{eqnarray}

For $q\ne t$, the QCD corrections to $\Gamma\left(H\to q\bar q\right)$ are
known up to ${\cal O}(\alpha_s^2)$.
In the $\overline{\rm MS}$ scheme, the result is \cite{bak}
\begin{eqnarray}
\label{hqqmsb}
\Gamma\left(H\to q\bar q\right)={3G_FM_Hm_q^2\over4\pi\sqrt2}
\left[\left(1-4{m_q^2\over M_H^2}\right)^{3/2}\right.&&\nonumber\\
{}+\left.
C_F{\alpha_s\over\pi}\left({17\over4}-30{m_q^2\over M_H^2}\right)
+K_2\left({\alpha_s\over\pi}\right)^2\right],&&
\end{eqnarray}
where $K_2\approx35.9399-1.3586N_F$ \cite{gor}, with $N_F$ being the number
of quark flavours active at scale $M_H$, and it is understood that $\alpha_s$
and $m_q$ are to be evaluated at $\mu=M_H$.
We note in passing that Eq.~(\ref{hqqmsb}) may be translated into the on-shell
scheme by using the above relation between $M_q$ and $m_q(M_H)$ \cite{kat}.
The difference between these two evaluations is extremely small \cite{kat},
which indicates that the residual uncertainty due to the lack of
knowledge of the ${\cal O}(\alpha_s^2m_q^2/M_H^2)$ and
${\cal O}(\alpha_s^3)$ terms is likely to be inconsequential for
practical purposes.

\subsection{Next-to-leading order corrections to $\Gamma(H\to gg)$}

The hadronic width of the Higgs boson receives contributions also from
the $H\to gg$ channel, which is mediated by massive-quark triangles,
and related higher-order processes.
The respective partial width is well approximated by \cite{djo}
\begin{eqnarray}
\Gamma\left(H\to gg(g),gq\bar q \right)&\n=\n&\Gamma(H\to gg)
\left[1+{\alpha_s\over\pi}\left({95\over4}\right.\right.\nonumber\\
&\n-\n&\left.\left.{7\over6}N_F\right)\right],
\end{eqnarray}
where $N_F$ is the number of quark flavours active at scale $M_H$, and
\cite{jel}
\begin{equation}
\Gamma(H\to gg)={\alpha_s^2G_FM_H^3\over36\pi^2\sqrt2}
\left(1+{7\over60}{M_H^2\over m_t^2}\right).
\end{equation}

\begin{figure}[h]
\centerline{\bf FIGURE CAPTIONS}
\caption{Universal electroweak correction factor for
$\Gamma\left(H\rightarrow f\bar f\,\right)$ to ${\cal O}(G_FM_H^2)$
and ${\cal O}(G_F^2M_H^4)$ as as a function of $M_H$.
In each order, the expanded result, Eq.~(\protect\ref{eqexp}), is compared
with the calculation where the one-particle-reducible Higgs-boson
self-energy diagrams are resummed, Eq.~(\protect\ref{eqres}).}
\label{figone}

\end{figure}

\end{document}